\documentclass[conference]{IEEEtran}
\IEEEoverridecommandlockouts  
\usepackage{amsmath,amssymb,epsfig,url,cite}
\usepackage{graphicx,footmisc}
\usepackage{algorithm}
\usepackage{algpseudocode}
\usepackage{siunitx}
\usepackage{tikz}
\usepackage{bbm}
\usepackage{subcaption}

\makeatletter
\def\BState{\State\hskip-\ALG@thistlm}
\makeatother
\newcommand{\T}{{T_1}}
\newcommand{\F}{{F_1}}
\newcommand{\Ts}{{T_2}}
\newcommand{\Fs}{{F_2}}
\newcommand\Algphase[1]{%
\Statex\hspace*{\dimexpr-\algorithmicindent--9pt}\textbf{#1}%
} 

\title{Spectrum Sharing for LTE-A Network in \\ TV White Space}
\author{
        \IEEEauthorblockN{Meghna Khaturia\IEEEauthorrefmark{1}, Sweety Suman\IEEEauthorrefmark{1}, Abhay Karandikar and Prasanna Chaporkar}
        \IEEEauthorblockA{Department of Electrical Engineering, Indian Institute of Technology Bombay, Mumbai-400076 \\ Email: \{meghnak,sweetysuman,karandi,chaporkar\}@ee.iitb.ac.in}
        \thanks{\IEEEauthorrefmark{1}{\vspace{-0.75cm}These authors have contributed equally to this work.}}
        \vspace{-1cm}
        }

\begin{document}
	
\maketitle
\IEEEpeerreviewmaketitle


\begin{abstract}

Rural areas in the developing countries are predominantly devoid of Internet access as it is not viable for operators to provide broadband service in these areas. To solve this problem, we propose a \textit{middle mile} Long Term Evolution Advanced (LTE-A) network operating in TV white space to connect villages to an optical Point of Presence (PoP) located in the vicinity of a rural area. We study the problem of spectrum sharing for the middle mile networks deployed by multiple operators. A graph theory based Fairness Constrained Channel Allocation (FCCA) algorithm is proposed, employing Carrier Aggregation (CA) and Listen Before Talk (LBT) features of LTE-A. We perform extensive system level simulations to demonstrate that FCCA not only increases spectral efficiency but also improves system fairness.

\end{abstract}


\section{Introduction}

\label{intro}
The world has seen a vast growth in communication technology and yet 52\% of the global population is still unconnected~\cite{unbb}, majority of which live in developing countries. The broadband penetration in the rural areas of developing countries is even worse due to high cost of infrastructure, difficult terrain, sparse population density and low Average Revenue per User (ARPU). A low cost broadband access to the end users in these areas can be provided by deploying Wi-Fi Access Points (APs). However, laying fiber to backhaul each and every Wi-Fi AP becomes infeasible as it is time consuming and expensive. If an optical Point of Presence (PoP) is located in the vicinity of a rural area, then a wireless \textit{middle mile network}, as proposed in~\cite{IEEEcommmag}, can be established to connect the optical PoP to the WiFi APs in the villages.

A possible solution for connecting the PoP to the APs is to use TV UHF band in the middle mile network as it is highly underutilized in many developing countries. In India, more than $100$~MHz of TV UHF band ($470$-$585$~MHz) is unused (commonly referred to as TV white space)~\cite{Gaurang}. Owing to the propagation characteristics of this band, it is possible to obtain large coverage area even with low power transmission. This will enable the use of renewable energy sources which is a highly desirable feature to develop an affordable technology for rural areas. Consequently, the operators can be encouraged to deploy a middle mile network in these areas. However, a middle mile network using the TV UHF band is not a plug and play solution. As multiple operators have to coexist in the same band, there will be huge interference among them which will lead to low spectral efficiency. Hence, it is important to design a spectrum sharing scheme which not only increases the spectral efficiency but also guarantees a fair share of spectrum to the operators. The major challenge in designing the above scheme is that it should be based on trivial information which can be easily availed from operators since they will be unwilling to share sensitive information. 

The above-mentioned spectrum sharing scenario is very similar to the limited spectrum pooling where a limited number of operators share a common pool of spectrum, obeying the rules of spectrum access set by multi-lateral agreements. Limited spectrum pooling has been studied in the context of heterogeneous networks in~\cite{smallcell} and \cite{dense}. In \cite{smallcell}, two operators pool equal bandwidth for sharing among small cells. An operator has preemptive priority over its own share of pooled spectrum. 
In~\cite{dense}, spectrum sharing is studied for a dense small cell network of two operators. In our system, the operators have equal priority over the spectrum in contrast to the pre-emptive nature of the scheme in~\cite{smallcell}. Moreover, in~\cite{smallcell,dense}, scheduling information has to be shared among operators which is infeasible in general.
In literature, game theoretic models are also employed to solve the problem of spectrum sharing. Two operators dynamically share the spectrum by playing a non-zero sum game in~\cite{gameKamal}. 
In~\cite{repeatedgames}, the authors model spectrum sharing among operators as a non-cooperative repeated game. The main concern in the above models is that it may result in inefficient Nash Equilibrium depending on the utility function selected by the operator. Moreover, the schemes discussed in \cite{gameKamal,repeatedgames} do not guarantee fairness and are also difficult to implement in realistic scenario. All the above literature discusses the spectrum sharing among only two operators. Generalization to multiple operators has not been studied in literature and is a challenging problem that we tackle in this paper. In this paper, we address the above issues while solving the spectrum sharing problem in rural setting.


The main contributions of this paper are:

\begin{itemize}
    \item Analysis of low power middle mile LTE-A network in TV UHF band with its coverage radius estimates.
    \item A graph theory based algorithm for spectrum sharing among multiple operators employing Listen Before Talk (LBT) and Carrier Aggregation (CA) features provided by LTE-A standard. 
    \item Performance evaluation of the proposed algorithm by system level simulations using Network Simulator-3 (ns3)~\cite{ns3}.
\end{itemize}


The organization of the rest of the paper is as follows. In Section~\ref{specalloc}, we discuss spectrum sharing among
multiple operators and also establish mathematical formulation for the same.
In Section~\ref{FCCA}, we propose a channel allocation algorithm to solve the spectrum sharing problem. In Section~\ref{res}, we analyse the performance of the proposed algorithm using ns-3 simulations. Finally, Section~\ref{conclu} concludes the paper.


\section{Spectrum Sharing Problem} 
\label{specalloc}
\subsection{Network Architecture}
\label{sec:sys}
\begin{figure}[!ht]
	
        \centering
        \includegraphics[scale=0.4 ]{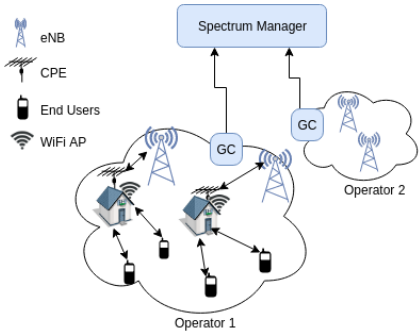}
        \caption{Overview of a multi-operator middle mile network}
        \label{fig:sys}
        \vspace{-0.5cm}
\end{figure}

We consider a middle mile LTE-A network operating in the TV UHF band. We assume that a portion of this band is available to multiple operators to deploy their networks in rural areas. This portion is divided into multiple orthogonal channels of equal bandwidth. The multi-operator middle mile network architecture is illustrated in Fig.~\ref{fig:sys}. The network comprises of a centralized entity called the \textit{Spectrum Manager (SM)} which is responsible for channel allocation to the operators. Multiple low power evolved NodeBs (eNBs) are deployed in a given area, preferably, in the vicinity of an optical PoP. Even though an eNB transmits at a very low power, the coverage area is typically large owing to the propagation characteristics of TV UHF band. Each operator has an entity called Gateway Controller (GC) which acts as an interface to communicate with the SM. Multiple LTE-A Customer Premise Equipments (CPEs) are served by each eNB. A CPE connects to one or many WiFi APs installed in a village. An end user accesses broadband services through a WiFi AP. We assume that the end users are uniformly distributed in a given area. 

The operator registers itself with the SM to access the TV UHF band. GC collects the topology details like antenna height, location and transmit power of each eNB under an operator and communicates it to the SM. No other details such as user scheduling information are shared to the SM for channel allocation. The SM then maintains a database of the information shared by GCs of all operators.
The SM treats all operators equally. As we have considered that the end users are uniformly distributed in a given area, the average throughput requirement at each eNB is equal. Hence, an operator gives equal priority to all its eNBs. Note that in further discussion we consider each eNB as an independent network entity. Henceforth, we study the spectrum sharing problem with respect to an eNB, irrespective of the operator. The channel allocated by the SM is communicated to an eNB of an operator through its GC.

For simplicity, we consider the \textit{Protocol Interference Model}~\cite{protocol} to model the interference between eNBs. In accordance with this model, two eNBs interfere with each other if they are operating on the same channel and the euclidean distance between them is less than a certain threshold distance. The protocol model formulates interference state as a binary symmetric matrix, where each element of the matrix indicates whether or not the two eNBs interfere with each other. 

\vspace{-0.1cm}
\subsection{System Model}
\label{sec:sysmod}
Consider a set $\mathcal{K} = \{1,2,...,K\}$ representing total number of eNBs belonging to all the operators in the network. Let $\mathcal{L}_{k} = \{1,2,...,L_{k}\}$ be the set of $L_{k}$ CPEs served by the eNB$_k\forall k\in\mathcal{K}$. The set of channels available at the SM is given by $\mathcal{M} = \{1,2,...,M\}$. Also, let $\mathcal{M}_{k}\subset \mathcal{M}$ be the set of channels assigned to eNB${_k}$. The eNB$_k$ allocates resources to its users from the assigned channels.

The SM allocates channel to the eNBs depending on the interference state of the network. Let $C = \{c_{k,j} | c_{k,j} \in \{0,1\}\}_{K\text{x}K}$ be a binary symmetric $K\text{x}K$ matrix where $(k,j)\in\mathcal{K}X\mathcal{K}$, represents the interference state such that $c_{k,j}=1$ when $\text{eNB}_k$ and $\text{eNB}_j$ interfere with each other, else $c_{k,j}=0$.
In addition to allocating channels, SM also defines the mode in which the channel has to be used. The mode of access can be \textit{shared} or \textit{dedicated}. If the mode of a channel assigned to an eNB is dedicated, then that channel does not get allocated to its neighbours. If the mode of access of the assigned channel is shared, then it is to be shared with the neighbours using some sharing mechanism. The channel allocation is given by the two matrices $A$ and $B$ which are defined as follows:
\begin{itemize}
        \item{\textit{Channel Allocation Matrix ($A$):}} We define channel allocation matrix as $A = \{a_{k,m}|a_{k,m}\in\{0,1\}\}_{K\text{x}M}$ where $k\in\mathcal{K}$ and $m\in\mathcal{M}$ such that $a_{k,m} = 1$, if channel $m$ is assigned to eNB$_k$, otherwise $a_{k,m} = 0$.
       
        \item{\textit{Mode Allocation Matrix ($B$):}} $B = \{b_{k,m} | b_{k,m} \in \{0,1\}\}_{K\text{x}M}$ is a $K\text{ by }M$ binary matrix where $k\in\mathcal{K}$ and $m\in\mathcal{M}$. $B$ represents the mode of access on the allocated channel such that $b_{k,m} = 1$, if allocated channel $a_{k,m}$ is to be shared, otherwise $b_{k,m} = 0$.
    \end{itemize}

SM assigns a single channel or multiple channels to an eNB. Carrier Aggregation (CA) feature is required at eNB for cross channel scheduling of resources~\cite{lte-ca}. 
When the mode of the allocated channel is \textit{shared}, Listen Before Talk (LBT) is used for sharing the channel. LBT is a mechanism in which a radio transmitter performs Clear Channel Assessment (CCA) to opportunistically transmit over an idle channel. 
The LBT mechanism has been discussed for the coexistence of LTE-A and Wi-Fi system~\cite{TS36.889}. We have used LBT for the coexistence among LTE-A systems in this work. 

Once the channel assignment is done by the SM, an eNB allocates resources from the assigned channels to its associated CPEs in a proportional fair manner. The sum throughput at eNB$_k$ is a function of $A$ and $B$ and is given by $T_k(A,B)$.
 We quantify the fairness $F$ of the system using Jain's Fairness Index (JFI) as below:
\begin{equation}
            \label{eqn:jfi}
           F = \frac{\left(\sum\limits_{k=1}^{K}T_k(A,B)\right)^2}{K\times\sum\limits_{k=1}^{K}T_k(A,B)^2}.
\end{equation}
\subsection{Problem Formulation}
The spectrum sharing problem can be modeled as a system throughput maximization problem under the fairness constraint. Mathematically, the problem can be stated as follows:
    \begin{equation}
                         \label{eq:obj}
                         (A^\star,B^\star) = \underset{A,B}{\arg\max} \left( \sum\limits_{k=1}^{K}T_{k}(A,B)\right), \\
                         \vspace{-0.3cm}
    \end{equation}
     
    \begin{align*}
    \text{subject to}   \quad F\geqslant \delta.
    \end{align*}
where $\delta$ is the constrained value of fairness.
There are two major challenges in obtaining an optimal solution of this problem. Firstly, this is a combinatorial optimization problem which is known to be NP-complete. 
Secondly, to determine an optimal solution, a closed form expression for throughput is required at the eNB. The mathematical expression for LBT throughput can be obtained only for a network which forms a complete graph. In our case the network graph is not complete. Therefore, in the following section, we propose a heuristic based graph theoretic algorithm to solve the above problem sub-optimally. 


\section{Fairness Constrained Channel Allocation (FCCA)}
\label{FCCA}
We first review the traditional graph coloring problem and then describe FCCA algorithm in detail.
The system can be modeled as a conflict graph $G(V,E)$, where $V$ represents set of all eNBs and $E$ denotes the set of edges. An edge between any two eNBs implies that the eNBs are interfering with each other i.e. $E := \{(k,j) | c_{k,j}=1,\; \forall k,j\in \mathcal{K}\}$ where $c_{k,j}$ is an element of the interference matrix, $C$ defined in Section~\ref{sec:sysmod}. In the traditional \textit{Graph Coloring} problem, colors are to be assigned to the vertices such that vertices with an edge between them do not get the same color.
The colors represent the available set of channels denoted by $\mathcal{M}$. Note that the numbers of colors i.e. the channels are considered to be fixed.
 
We now exploit the above graph coloring technique in our algorithm. The FCCA algorithm takes graph $G$ as an input and outputs the allocation matrices, $A$ and $B$. Here, $G$ is the graph representing the network as discussed above. In this method, the channels are assigned to the eNBs according to two sub-algorithms which are described next.
\begin{enumerate}
    \item \textit{Multiple Dedicated Channel Allocation (MDCA)}: In this sub-algorithm, multiple dedicated channels are assigned to an eNB by using greedy graph-coloring method iteratively. It is possible to assign multiple channels to an eNB if the total number of neighbours of an eNB is less than the total number of channels.
    \item \textit{One Dedicated Rest Shared Channel Allocation (ODRS-CA)}: In this sub-algorithm, the channel assignment is done in two steps. In the first step, a single dedicated channel is assigned to each eNB. 
    Then, the set $\mathcal{N}_k$, containing all the channels which are not assigned to the neighbours of eNB$_k$ is obtained. In the second step, all the channels contained in $\mathcal{N}_k$ are assigned to eNB$_k$ in shared mode.
    
\end{enumerate}
For a given network topology, the output of the above mentioned sub-algorithms are compared to decide the final channel allocation as described in Algorithm~\ref{alg}. There is always a guarantee that at least one channel will be allocated to each eNB irrespective of which sub-algorithm is chosen. Hence, a certain level of fairness is always ensured. Ideally, the value of $\delta$ should be equal to $1$ for complete fairness. However, if we give more preference to the fairness, the system throughput will be compromised. Hence, we choose the above $\delta$ equal to $0.75$ to strike a balance between throughput and fairness.

\begin{algorithm}
\caption{Fairness Constrained Channel Allocation}  
\label{alg}
\begin{algorithmic}

\Require {Graph G}
\State {$\delta = 0.75$}
\Algphase{Sub-Algorithm 1 : MDCA}
\While{$\mathcal{N}_k$ is non empty for all $K$}
\For{each $k$ from 1 to $K$}
    \State find $\mathcal{E}_k$, set of channels assigned to neighbours of $k$
    \State obtain $\mathcal{N}_k=\{\mathcal{M}\}\setminus \{\mathcal{E}_k\}$, set of feasible channels for eNB$_k$,
    \State $q \gets \min\mathcal{N}_k$, ${a_{k,q}} \gets {1}$, ${b_{k,q}} \gets {0}$
\EndFor
\EndWhile
\State ${\T} \gets {T(A, B)}$, ${\F} \gets {F(A, B)}$, $A_1 \gets A$, $B_1\gets B$

\Algphase{Sub-Algorithm 2 : ODRS-CA}
\For{each $k$ from 1 to ${K}$}

    \State find $\mathcal{E}_k$
    \State obtain $\mathcal{N}_k=\{\mathcal{M}\}\setminus \{\mathcal{E}_k\}$ 
    \State $q \gets \min\mathcal{N}_k$, ${a_{k,q}} \gets {1}$, ${b_{k,q}} \gets {0}$

\EndFor

\For{each $k$ from 1 to ${K}$}
    \State find $\mathcal{E}_k $
    \State obtain $\mathcal{N}_k=\{\mathcal{M}\}\setminus \{\mathcal{E}_k\}$ 
    \State ${a_{k,q}} \gets {1}$, ${b_{k,q}} \gets {1} \quad \forall q\in\mathcal{N}_k$

\EndFor

\State ${\Ts} \gets {T(A, B)}$, ${\Fs} \gets {F(A, B)}$, $A_2 \gets A$, $B_2\gets B$ 
\Algphase{Result:}
Check $\F$ and $\Fs$ and choose ($A^\star,B^\star$) such that the fairness is greater than $\delta$. If both are greater than $\delta$ then choose ($A^\star,B^\star$) corresponding to $\max(\T,\Ts)$. 

\Algphase{return $A^\ast$, $B^\ast$ }

\end{algorithmic}
\end{algorithm}

\vspace{-0.15cm}

\section{Performance Analysis}
\label{res}
In this section, we present the results of ns-3 simulations to assess the performance of FCCA algorithm. We also compare the proposed approach with few other coexistence approaches.

\vspace{-0.1cm}
\subsection{Scenario Description}
\begin{figure}[!ht]
	 
	\centering    
	\includegraphics[scale=0.58]{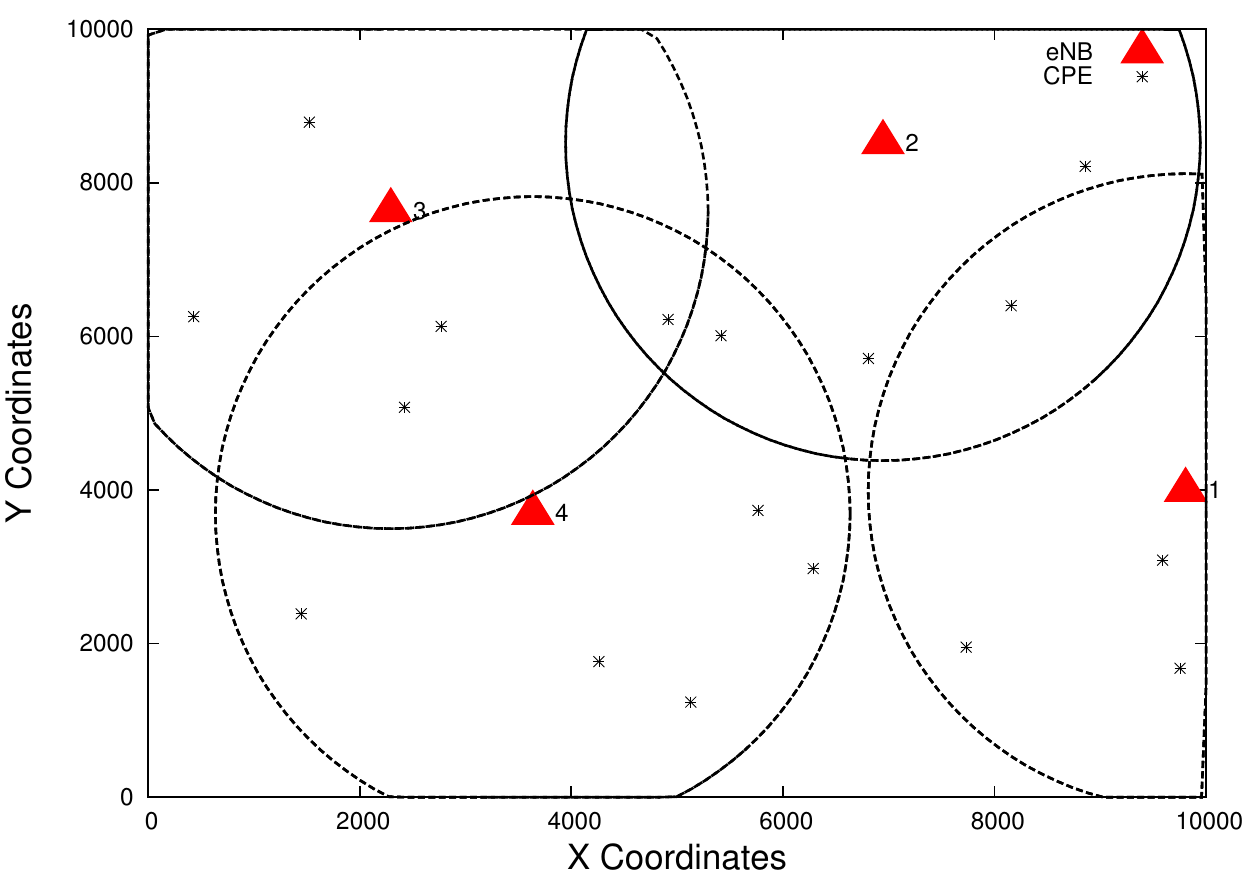}
    \caption{An example topology of the network.}
    \label{fig:coordinate}
   	\vspace{-0.4cm}
\end{figure}

We assume that $20$~MHz of the TV UHF band is available for middle mile network. This band is further divided into $4$ orthogonal channels of $5$~MHz each. All channels are assumed to be identical. 
As the rural areas are sparsely populated, we assume that an eNB will not get interference from three or more eNBs. The eNBs are deployed uniformly at random in an area of $100~\text{km}^2$ as shown in Fig.~\ref{fig:coordinate}. The CPEs are distributed uniformly within the coverage area of an eNB. Each eNB is assumed to serve $5$ stationary CPEs. For constructing the conflict graph using protocol interference model, we consider a distance of $4$~km between eNBs. If the distance between eNBs is less than $4$~km, then they interfere with each other. We perform ns-3 simulations over $100$ random topologies. All the performance metrics are averaged over such realizations. 
The simulation parameters are given in Table~\ref{PhyParam}.
With a static rural environment along with stationary CPEs, it is reasonable to rule out fast fading effects in our scenario. We consider only saturated downlink transmission in this work i.e. at each eNB, saturated traffic is generated for each of the associated CPEs. 

\begin{table}
        \centering
        \caption{Simulation Parameters}
        \label{PhyParam}
        \renewcommand{\arraystretch}{0.9}
        \begin{tabular}{|l|r|}
            \hline
            \textbf{Parameters}                             & \textbf{Values}   \\ \hline \hline
            Central Frequency($f_c$)                                  & $500$-$520$ MHz         \\ \hline   
            Transmit Power ($P_t$)                                & $18$ dBm             \\ \hline
            Receiver Sensitivity ($RS$)						    & $-101$ dBm ~\cite{Rxsense}\\ \hline 
            Cable Loss ($CL$)								& $2$ dB				\\ \hline
            Receiver Noise Figure	($NF$)					& $7$ dB				\\ \hline
            Transmitter Antenna Gain  ($G_t$)                  & $10$ dB               \\ \hline
            Receiver Antenna Gain   ($G_r$)                 & $0$ dB               \\ \hline
            Transmitter Antenna Height  ($h_t$)                & $30$ m              \\ \hline
            Receiver Antenna Height  ($h_r$)                & $5$ m               \\ \hline
            Slot Time                                       & $9~\mu s$         \\ \hline
            Transmit Opportunity ($TxOp$)                   & $10$~ms           \\ \hline
            Detection Threshold                                             & $-62$~dBm        \\ \hline
            Simulation time                                 & $30$ s             \\ \hline
        \end{tabular}
\vspace{-0.3cm}  
    \end{table}

\vspace{-0.1cm}
\subsection{Coverage Radius of eNB operating in TV UHF band}
\label{cov-area}
The coverage radius of a transmitter is defined as the maximum allowed distance between the transmitter and the receiver such that they can communicate. We calculate the coverage radius of an eNB using the equation:
\begin{equation}
    \label{eqn:coverage}
  RS = P_t + G_t + G_r - PL(d,h_t,h_r,f_c) - CL - NF,
 \end{equation}
where $RS$, $P_t$, $G_t$, $G_r$, $CL$, $NF$, $h_t$, $h_r$ and $f_c$ are as per the Table~\ref{PhyParam}.
$PL(d,h_t,h_r,f_c)$ is the path loss which is also a function of distance $d$ between transmitter and receiver. Hata model for Suburban Areas is used to calculate path loss~\cite{hata}. The transmit power, $P_t$, of an eNB is $18$~dBm which is significantly low.
For the values of the parameters given in Table~\ref{PhyParam}, the coverage radius of eNB comes out to be approximately $3$~km.
\vspace{-0.1cm}    

\vspace{-0.1cm}
\subsection{Results}

\begin{figure*}[!ht]
    \centering
    \begin{subfigure}[b]{0.4\textwidth}
    	\fbox{\includegraphics[width=\textwidth]{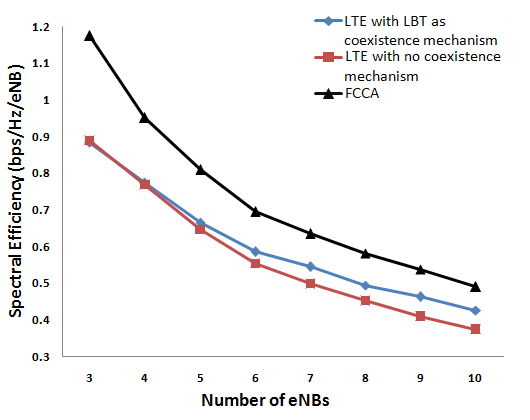}}     
    	\caption{Spectral Efficiency of eNB vs. number of eNBs deployed in $100~\text{km}^2$ area}  
        \label{speff}
    \end{subfigure}
    \quad \quad \quad \quad \quad \quad \quad
    \begin{subfigure}[b]{0.4\textwidth}
        \fbox{\includegraphics[width=\textwidth]{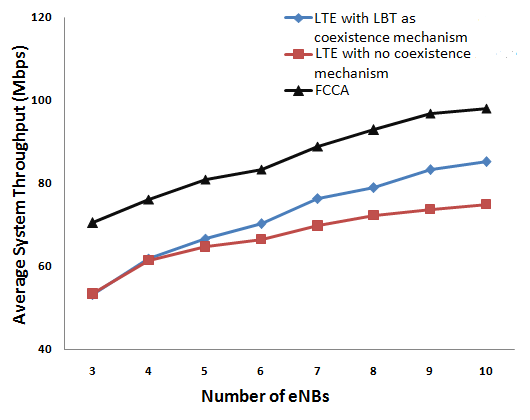}}
        \caption{Average throughput per eNB vs. number of eNBs deployed in $100~\text{km}^2$ area}
        \label{thrpt }
    \end{subfigure}
    \caption{Comparative analysis of average system throughput and spectral efficiency for the three schemes.}
     \label{fig:res}
    \vspace{-0.25cm}
\end{figure*}

\begin{figure}
	
    \centering
    \fbox{\includegraphics[width=0.4\textwidth]{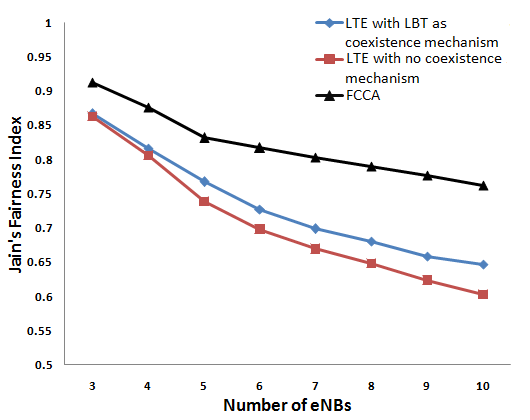}}
     \caption{An example topology of the network.}
    \label{JFI}
    \vspace{-0.5cm}
\end{figure}

We analyze three performance metrics to assess the performance of the proposed FCCA algorithm: a) \textit{Spectral Efficiency} b) \textit{Average System Throughput per eNB} and c) \textit{Jain's Fairness Index}. The performance of these metrics are observed with respect to an increase in the network density i.e. we increase the number of eNBs from $3$ to $10$ in a fixed area of $100$~km$^2$. 
In our results we present spectral efficiency per eNB which is measured in bits/s/Hz. We have used Jain's Fairness Index to quantify how fairly the available band is shared among the eNBs. 

In Fig.~\ref{fig:res}, we compare the spectral efficiency and the average system throughout of FCCA with two other schemes i) LTE with no-coexistence mechanism ii) LTE with LBT as the coexistence mechanism. Clearly, the FCCA algorithm outperforms both the schemes in both the metrics. In the first scheme, the entire $20$~MHz band is used by all eNBs without any coexistence mechanism. Here the spectral efficiency is poor due to interference among the eNBs. In the second scheme, the entire $20$~MHz band is shared among all the eNBs using LBT. In this case, the performance is poor as the transmission time is wasted in contention. The FCCA algorithm performs better than the above schemes as it takes the topology into consideration for allocating the channels. As shown in Fig.~\ref{JFI}, the fairness of the FCCA algorithm is also better than the other two schemes. The proposed algorithm guarantees an excellent fairness index of $0.76$ even in the case $10$ eNBs per $100~\text{km}^2$.

It is important to analyse the performance of the proposed algorithm under the best case and the worst case scenario. Consider a very sparse network (best case) where only three eNBs are deployed in the given area. When $100$ random topologies are simulated under this setting, it is observed that in $50$\% of the cases MDCA is preferred and in the remaining cases ODRS-CA is preferred. This result highlights the fact that orthogonal channel allocation may not always give the best result. In a very dense network (worst case) where $10$ eNBs are deployed in the given area, in 61\% cases ODRS-CA is preferred between the two sub-algorithm. This result further emphasizes the use of LBT as sharing mechanism for better system performance in terms of spectral efficiency and fairness.

\subsection{Average Throughput vs Demand}
We now consider a typical rural setting in India. An optical PoP is present at the village office called Gram Panchayat (GP) which typically serves 2 villages. Approximately 5 GPs are present in an area of $100$~km$^2$ serving 10 villages. The average population of a village in India is $1000$. There is one subscriber per household i.e. one person in a house of $5$ will subscribe to broadband service. Consider a minimum broadband throughput of $2$~Mbps with the contention ratio of $1:50$. Thus, the average throughput requirement under the above scenario is $(1000$~people$\times10$~villages$\times2$~Mbps)$/(50\times5)=80$~Mbps. If 5 eNBs are deployed, each at one GP, the average throughput demand that can be served using FCCA algorithm is approximately $83$~Mbps. Hence, the average throughput requirement of the above setting can be easily met. 
\vspace{-0.3cm}
\section{Conclusion}
\label{conclu}
We have discussed the problem of poor broadband penetration in rural areas of developing countries. To solve this problem, we have proposed a middle mile LTE-A network operating in TV UHF band. We have presented a centralized graph theory based channel allocation algorithm with a novel concept of allocating a combination of shared and dedicated channel to an eNB. The performance of the algorithm has been studied using ns-3 simulations. The results demonstrate that it increases both the spectral efficiency and the fairness among operators in a network. We have also compared the obtained average throughput with the throughput demand generated by a rural setting in India. We note that the proposed scheme easily meets the throughput demand in a rural area.

\end{document}